\documentclass[12pt]{article}
\begin{document}

\title{The Power Spectrum for a Multi-Component Inflaton \\
to Second-Order Corrections in the Slow-Roll Expansion}

\author{
Jin-Ook Gong\footnote{devourer@muon.kaist.ac.kr}
\hspace{2cm}
Ewan D. Stewart\footnote{ewan@kaist.ac.kr} \\
{\em Department of Physics, KAIST, Daejeon 305-701, South Korea}}
\date{\today}
\maketitle

\begin{abstract}
We derive the power spectrum $\mathcal P(k)$ of the density
perturbations produced during inflation up to second-order
corrections in the standard slow-roll approximation for an
inflaton with more than one degree of freedom. We also present the
spectral index $n$ up to first-order corrections including
previously missing terms, and the running ${dn}/{d\ln k}$ to
leading order.
\end{abstract}

\vspace*{-65ex}
\hspace*{\fill} KAIST-TH 2002/01

\thispagestyle{empty}
\setcounter{page}{0}
\newpage
\setcounter{page}{1}

\section{Introduction}

Inflation \cite{iu} generates the primordial perturbations which
are the origin of the rich structures, such as galaxies and
clusters of galaxy clusters, observed in the universe today. The
power spectrum of these perturbations is constrained to be
approximately scale invariant by a combination of the measurements
of anisotropies on large and small angular scales, as well as
galaxy surveys \cite{ls+ss}. Moreover, forthcoming measurements
such as the MAP and Planck satellites and the Sloan Digital Sky
Survey will probe the power spectrum with greater accuracy,
lowering the observational errors further. Planck, for example, is
expected to be able to measure the power spectrum of the cosmic
microwave background with an error of a few percent \cite{psm}.
Thus, it is very important to calculate the power spectrum
precisely, so that we can make use of the observations fully.

The calculation of the density perturbations for an inflaton with
a single degree of freedom has been studied for a long time
\cite{oldps,sl,sg}, and there have been extensive works on the
multi-component inflaton case, especially recently
\cite{oldmulti,ss,ns,r-multi}. Among these results, those with
first-order corrections \cite{sl,ns} are expected to have errors
of the order of one percent or smaller for the power spectrum
\cite{pt}, which is comparable with the expected errors in the
planned observations. Thus we should reduce the errors and improve
the accuracy of the calculations. However, the only existing
result including second-order corrections \cite{sg} considered
only the single-component case. Therefore, it is needed to
consider the multi-component inflaton with higher order
corrections.

In this paper, we extend our previous formalism \cite{sg} to the
multi-component case, and calculate the power spectrum up to
second-order corrections in the standard slow-roll
expansion\footnote{For a more general slow-roll expansion, see
Ref.~\cite{ds}.}.  We also present the spectral index $n$ up to
first-order corrections with new terms inconsistently neglected in
the previous result \cite{ns}, and the running $dn/d \ln k$ to
leading order for the first time. It is straightforward to check
that our results reduce to those of Ref.~\cite{sg} in the single
component case.

\section{The Spectrum for a Multi-Component Inflaton}

The power spectrum $\mathcal{P}(k)$ is defined by
\begin{equation}\label{ps}
\frac{2\pi^2}{k^3}\,\mathcal{P}(k)\,\delta^{(3)}(\mathbf{k-l}) =
\langle \mathcal{R}_\mathrm{c}(\mathbf{k})\,
{\mathcal{R}_\mathrm{c}}^\dagger(\mathbf{l}) \rangle \,,
\end{equation}
where $\langle \ \rangle$ denotes the vacuum expectation value and
$\mathcal{R}_\mathrm{c}$ is the intrinsic curvature perturbation
of the comoving hypersurfaces. In the single-component case, the
standard result for the power spectrum of the curvature
perturbation in the slow-roll approximation is
\begin{equation}\label{stdps}
\mathcal{P}(k) \simeq \left( \frac{H}{2\pi} \right)^2 \left(
\frac{H}{\dot\phi} \right)^2
\,.\end{equation}

The multi-component case is more subtle. Detailed arguments are
given in \cite{ss}; here we just give the principal idea. Let
$N=\int H dt$ be the number of $e$-folds of expansion. Then, on
super-horizon scales, we obtain \cite{ss}
\begin{equation}
\Delta\mathcal{R} = \delta N \,.
\end{equation}
That is, the change in the curvature $\mathcal{R}$ between the
initial and final hypersurfaces is given by the perturbation in
the number of $e$-folds of expansion. In particular, if we choose
the initial hypersurface to be flat, \mbox{i.e.}
$\mathcal{R}(t_1)=0$, and the final one to be comoving, we get
\begin{equation}
\mathcal{R}_\mathrm{c}(t_2) = \delta N \,.
\end{equation}
We take $t_1$ to be some time during slow-roll inflation, and
$t_2$ to be some time after inflation when
$\mathcal{R}_\mathrm{c}$ has become constant. The relevant scale
is assumed to be still outside the horizon at $t_2$. Now, $N$
depends on both $\phi(t_1)$ and $\dot\phi^\mathbf{a}(t_1)$, but as
$t_1$ is during slow-roll inflation, it will be convenient to
express this dependence in terms of $\phi(t_1)$ and
$\dot\phi_\perp^\mathbf{a}(t_1)$, where $\dot\phi_\perp^\mathbf{a}
\equiv \dot\phi^\mathbf{a} - \dot\phi_\mathrm{sr}^\mathbf{a}$ is
the deviation of $\dot\phi^\mathbf{a}$ from its value
$\dot\phi_\mathrm{sr}^\mathbf{a}$ on the exact slow-roll
trajectory passing through $\phi$. Then we can write\footnote{We
use the abstract index notation \cite{grbook}. A boldface
superscript index denotes a vector in the scalar field space, a
boldface subscript index denotes a covector, contractions are
denoted by repeated indices, and the metric $h_\mathbf{ab}$ and
its inverse $h^\mathbf{ab}$ are used to lower and raise indices,
respectively.}
\begin{equation}
\delta N
= \frac{\partial N}{\partial \phi^\mathbf{a}}\, \delta\phi^\mathbf{a}(t_1)
+ \frac{\partial N}{\partial \dot\phi_\perp^\mathbf{a}}\, \delta\dot\phi_\perp^\mathbf{a}(t_1)
\,.
\end{equation}
However, $\delta\dot\phi_\perp^\mathbf{a}$ is negligible on
super-horizon scales during slow-roll inflation. Therefore we get
\begin{equation}
\mathcal{R}_\mathrm{c}(t_2) = \delta N
= \frac{\partial N}{\partial \phi^\mathbf{a}}\, \delta\phi^\mathbf{a}(t_1)
\,.\end{equation}
Using Fourier expansion and substituting into Eq.~(\ref{ps}),
we obtain the power spectrum for a multi-component inflaton
\begin{equation}\label{p1}
\frac{2\pi^2}{k^3}\,\mathcal{P}(k)\,\delta^{(3)}(\mathbf{k-l})
= \frac{\partial N}{\partial\phi^\mathbf{a}}
\frac{\partial N}{\partial\phi^\mathbf{b}}
\left\langle \delta\phi^\mathbf{a}(\mathbf{k})\,
{\delta\phi^\mathbf{b}}^\dagger(\mathbf{l}) \right\rangle
\,.\end{equation}

Now, to leading order in the slow-roll approximation, \cite{ss}
\begin{equation}
\left\langle \delta\phi^\mathbf{a}(\mathbf{k})\,
{\delta\phi^\mathbf{b}}^\dagger(\mathbf{l}) \right\rangle
\simeq \left.
\frac{H^2}{2k^3}\,\delta^{(3)}(\mathbf{k-l})\,h^\mathbf{ab}
\right|_{aH=k}
\,.\end{equation}
Therefore
\begin{equation}
\mathcal{P}(k) \simeq \left. \left(\frac{H}{2\pi}\right)^2
h^\mathbf{ab}
\frac{\partial N}{\partial\phi^\mathbf{a}}
\frac{\partial N}{\partial\phi^\mathbf{b}}
\right|_{aH=k}
\,.\end{equation}
Defining $e^N_\mathbf{a}$ to be the unit covector in the direction
$\partial N / \partial\phi^\mathbf{a}$ and using
\begin{equation}
\dot\phi^\mathbf{a} \frac{\partial N}{\partial\phi^\mathbf{a}}
= - H
\end{equation}
we can write this as
\begin{equation}
\mathcal{P}(k) \simeq \left. \left(\frac{H}{2\pi}\right)^2 \left(
\frac{H}{e^N_\mathbf{a} \dot\phi^\mathbf{a}} \right)^2
\right|_{aH=k} \,,
\end{equation}
and so in the single-component case we recover Eq.~(\ref{stdps}),
but with some greater insight into its meaning.

\section{The Calculation}

The action during inflation is
\begin{equation}\label{action}
S = \int d^4 x \sqrt{-g} \left[ - \frac{1}{2} R + \frac{1}{2}
h_\mathbf{ab} g^{\mu\nu} (\partial_\mu \phi)^\mathbf{a}
(\partial_\nu \phi)^\mathbf{b} - V(\phi) \right] \,,
\end{equation}
where $g_{\mu\nu}$ and $R$ are the metric and curvature scalar in
the spacetime, and $h_\mathbf{ab}$ is the metric in the scalar
field space. The equation of motion for $\phi$ in the spatially
flat homogeneous and isotropic background is
\begin{equation}\label{bgeq}
\ddot\phi^\mathbf{a} + 3H\dot\phi^\mathbf{a} + V^\mathbf{,a} = 0
\,,
\end{equation}
where $\ddot\phi^\mathbf{a} \equiv D\dot\phi^\mathbf{a}/dt \equiv
\dot\phi^\mathbf{b}\nabla_\mathbf{b}\dot\phi^\mathbf{a}$ and
$\nabla_\mathbf{a}$ is the covariant derivative in the scalar
field space. We also have
\begin{equation}\label{etceq}
3H^2 = V + \frac{1}{2}\dot\phi^\mathbf{a}\dot\phi_\mathbf{a}
\hspace{2em} \mbox{and} \hspace{2em}
\dot{H} = -\frac{1}{2}\dot\phi^\mathbf{a}\dot\phi_\mathbf{a}
\,.\end{equation}

We define the slow-roll parameters
\begin{equation}\label{srp}
\epsilon \equiv \frac{1}{2}\frac{|\dot\phi|^2}{H^2}
\hspace{2em} \mbox{and} \hspace{2em}
\delta \equiv
\frac{\dot\phi_\mathbf{a} \ddot\phi^\mathbf{a}}{H |\dot\phi|^2}
\,,\end{equation}
and make the slow-roll assumptions
\begin{equation}\label{sra}
\epsilon = \mathcal{O}(\xi)
\hspace{2em} \mbox{and} \hspace{2em}
\delta = \mathcal{O}(\xi)
\,,\end{equation}
for some small parameter $\xi$.
We then have
\begin{equation}
\frac{\dot\epsilon}{H} = 2 \epsilon ( \epsilon + \delta ) = \mathcal{O}(\xi^2)
\end{equation}
and we will make the standard extra assumption\footnote{ For a
more general slow-roll approximation which doesn't make this extra
assumption, see Ref.~\cite{ds}.}
\begin{equation}\label{ssra}
\frac{\dot\delta}{H} = \mathcal{O}(\xi^2)
\,.\end{equation}

\subsection{Equation of motion for the perturbations}

The equation of motion for the inflaton perturbation
$\delta\phi^\mathbf{a}(\mathbf{k}, t)$ on flat hypersurfaces is
\cite{ss}
\begin{equation}\label{eqm}
\frac{D^2\delta\phi^\mathbf{a}}{dt^2}
+ 3H\frac{D\delta\phi^\mathbf{a}}{dt}
- R^\mathbf{a}_\mathbf{\ cdb} \dot\phi^\mathbf{c} \dot\phi^\mathbf{d}\,\delta\phi^\mathbf{b}
+ \left(\frac{k}{a}\right)^2 \delta\phi^\mathbf{a}
+ V^\mathbf{;ab} \delta\phi_\mathbf{b}
= \frac{1}{a^3}\frac{D}{dt} \left(
\frac{a^3}{H} \dot\phi^\mathbf{a} \dot\phi^\mathbf{b} \right)
\delta\phi_\mathbf{b}
\,,\end{equation}
where $R^\mathbf{a}_\mathbf{\ bcd}$ is the curvature tensor in the
scalar field space, defined by
\begin{equation}
R^\mathbf{a}_\mathbf{\ bcd} v^\mathbf{b}
\equiv
\left( \nabla_\mathbf{c}\nabla_\mathbf{d}
- \nabla_\mathbf{d}\nabla_\mathbf{c} \right) v^\mathbf{a}
\,.\end{equation}
Defining $\varphi^\mathbf{a} \equiv a\,\delta\phi^\mathbf{a}$,
the conformal time $d\eta \equiv dt/a$ and $x \equiv -k\eta$ we get
\begin{equation}\label{eqm'}
\frac{D^2\varphi^\mathbf{a}}{dx^2}
+ \left[ 1 - 2\left(\frac{a H}{k}\right)^2 \right] \varphi^\mathbf{a}
= \left(\frac{a H}{k}\right)^2 \left[
\frac{\dot{H}}{H^2} h^\mathbf{ab}
+ R^\mathbf{a\ \ b}_\mathbf{\ cd}
\frac{\dot\phi^\mathbf{c}}{H} \frac{\dot\phi^\mathbf{d}}{H}
- \frac{V^\mathbf{;ab}}{H^2}
+ \frac{1}{a^3H} \frac{D}{Hdt} \left( a^3H \frac{\dot\phi^\mathbf{a}}{H}\frac{\dot\phi^\mathbf{b}}{H} \right)
\right] \varphi_\mathbf{b}
\,.\end{equation}
Using
\begin{equation}\label{eta}
x = - k \int \frac{dt}{a}
= \frac{k}{aH} \left[ 1 + \epsilon + 3\epsilon^2
+ 2\epsilon\delta + \mathcal{O}\left(\xi^3\right) \right]
\end{equation}
the equation of motion is
\begin{equation}\label{feqm}
\frac{D^2\varphi^\mathbf{a}}{dx^2} + \left( 1 - \frac{2}{x^2}
\right) \varphi^\mathbf{a} = \frac{3}{x^2}
\zeta^\mathbf{a}_\mathbf{\ b} \varphi^\mathbf{b}
\end{equation}
where\footnote{$\zeta^\mathbf{ab}$ reduces to
$\epsilon^\mathbf{ab}$ of Ref.~\cite{ns} at order $\xi$, and to
$g/3$ of Ref.~\cite{sg} in the case of a single component
inflaton.}
\begin{eqnarray}\label{g}
\zeta^\mathbf{ab} & = & \left( \epsilon + 4\epsilon^2 +
\frac{8}{3}\epsilon\delta \right) h^\mathbf{ab} + \left( 1 +
\frac{5}{3} \epsilon \right)
\frac{\dot\phi^\mathbf{a}}{H}\frac{\dot\phi^\mathbf{b}}{H} +
\frac{1}{3} \frac{D}{H dt} \left(
\frac{\dot\phi^\mathbf{a}}{H}\frac{\dot\phi^\mathbf{b}}{H} \right)
\nonumber \\ & & \mbox{} + \frac{1}{3} \left( 1 + 2\epsilon
\right) R^\mathbf{a\ \ b}_\mathbf{\ cd}
\frac{\dot\phi^\mathbf{c}}{H}\frac{\dot\phi^\mathbf{d}}{H} -
\left( 1 + 2\epsilon \right) \frac{V^\mathbf{;ab}}{3H^2} +
\mathcal{O}\left(\xi^3\right) \,.\end{eqnarray}

\subsection{Quantization and Green's function solution}

The quantization conditions are
\begin{equation}
\left[ \varphi^\mathbf{a}(\mathbf{x},\eta),
\frac{D\varphi^\mathbf{b}}{\partial\eta}(\mathbf{y},\eta) \right]
= i\,\delta^{(3)}(\mathbf{x-y})\,h^\mathbf{ab}
\,,\end{equation}
otherwise zero.
With the Fourier transformation
\begin{equation}\label{fxform}
\varphi^\mathbf{a}(\mathbf{x},\eta)
= \int\frac{d^3k}{(2\pi)^{3/2}}\,
\varphi^\mathbf{a}(\mathbf{k},\eta)\,e^{i \mathbf{k \cdot x}}
\,,\end{equation}
the quantization conditions for the Fourier modes are
\begin{equation}
\left[
\varphi^\mathbf{a}(\mathbf{k},\eta),
\frac{D{\varphi^\mathbf{b}}^\dagger}{\partial\eta}(\mathbf{l},\eta)
\right]
= i\,\delta^{(3)}(\mathbf{k-l})\,h^\mathbf{ab}
\,,\end{equation}
otherwise zero.

Now, the solution of the homogeneous part of Eq.~(\ref{feqm}),
\begin{equation}\label{heq}
\frac{D^2\varphi_0^\mathbf{\ a}}{dx^2} + \left( 1 - \frac{2}{x^2}
\right) \varphi_0^\mathbf{\ a} = 0 \,,
\end{equation}
with the vacuum boundary condition at $x\rightarrow\infty$ is
\begin{equation}\label{hsol}
\varphi_0^\mathbf{\ a}(\mathbf{k},x) = \frac{1}{\sqrt{2k}} \left[
a^\mathbf{a}(\mathbf{k})\,\varphi_0(x) + a^{\dagger\mathbf{a}}
(\mathbf{-k})\,\varphi_0^*(x) \right] \,,
\end{equation}
where
\begin{equation}
\left[ a^\mathbf{a}(\mathbf{k}),
a^{\dagger\mathbf{b}}(\mathbf{l}) \right]
= \delta^{(3)}(\mathbf{k-l})\,h^\mathbf{ab}
\end{equation}
and
\begin{equation}
\varphi_0(x) = \left( 1 + \frac{i}{x} \right) e^{ix}
\,.\end{equation}
The Green's function solution of Eq.~(\ref{feqm}) with these
boundary conditions is
\begin{equation}\label{gfsol}
\varphi^\mathbf{a}(\mathbf{k}, x) = \varphi_0^\mathbf{\
a}(\mathbf{k}, x) + \frac{3i}{2} \int_x^{\infty} \frac{du}{u^2}\,
\zeta^\mathbf{a}_\mathbf{\ b}(k,u)\,
\varphi^\mathbf{b}(\mathbf{k}, u) \left[
\varphi_0^*(u)\,\varphi_0(x) - \varphi_0^*(x)\,\varphi_0(u)
\right] \,.\end{equation}

\subsection{Slow-roll expansion}

To implement the standard\footnote{ For a more general slow-roll
expansion, see Ref.~\cite{ds}.} slow-roll expansion, we expand
$\zeta^\mathbf{ab}$ in a power series in $\ln x$,
\begin{equation}\label{exp-g}
\zeta^\mathbf{ab} = \sum_{n=0}^{\infty} \zeta_{n+1}^\mathbf{\
ab}(k)\, \frac{(\ln x)^n}{n!} \,,
\end{equation}
and, in addition to Eqs.~(\ref{sra}) and~(\ref{ssra}), assume
\begin{equation}
{\zeta_n}^\mathbf{ab} = \mathcal{O}\left(\xi^n\right) \,.
\end{equation}
Therefore, to obtain the solution up to second-order corrections,
it is sufficient to consider
\begin{equation}\label{gexp}
\zeta^\mathbf{ab}={\zeta_1}^\mathbf{ab} + {\zeta_2}^\mathbf{ab}\ln
x \,,\end{equation} where
\begin{eqnarray}\label{g1}
{\zeta_1}^\mathbf{ab} & = & \left. \zeta^\mathbf{ab} \right|_{x=1}
= \left. \zeta^\mathbf{ab} \right|_{aH=k}
+ \mathcal{O}\left(\xi^3\right) \\
& = & \left( 1 + 4\epsilon + \frac{8}{3}\delta \right)
\epsilon h^\mathbf{ab}
+ \left( 1 + \frac{5}{3} \epsilon \right)
\frac{\dot\phi^\mathbf{a}}{H}\frac{\dot\phi^\mathbf{b}}{H}
+ \frac{1}{3} \frac{D}{H dt} \left( \frac{\dot\phi^\mathbf{a}}{H}\frac{\dot\phi^\mathbf{b}}{H} \right)
\nonumber \\ & & \left. \mbox{}
+ \frac{1}{3} \left( 1 + 2\epsilon \right)
R^\mathbf{a\ \ b}_\mathbf{\ cd}
\frac{\dot\phi^\mathbf{c}}{H}\frac{\dot\phi^\mathbf{d}}{H}
- \left( 1 + 2\epsilon \right) \frac{V^\mathbf{;ab}}{3H^2}
+ \mathcal{O}\left(\xi^3\right) \right|_{aH=k}
\end{eqnarray}
and
\begin{eqnarray}\label{g2}
{\zeta_2}^\mathbf{ab} & = & \left.\frac{D \zeta^\mathbf{ab}}{d \ln
x}\right|_{x=1} = - \left.\frac{D \zeta^\mathbf{ab}}{H
dt}\right|_{aH=k}
+ \mathcal{O}\left(\xi^3\right) \\
& = &
- 2 \epsilon \left( \epsilon + \delta \right) h^\mathbf{ab}
- \frac{D}{H dt} \left( \frac{\dot\phi^\mathbf{a}}{H}\frac{\dot\phi^\mathbf{b}}{H} \right)
\nonumber \\ & & \left. \mbox{}
- \frac{1}{3}\frac{D}{H dt}
\left( R^\mathbf{a\ \ b}_\mathbf{\ cd}
\frac{\dot\phi^\mathbf{c}}{H}\frac{\dot\phi^\mathbf{d}}{H} \right)
+ \frac{D}{H dt} \left( \frac{V^{;\bf ab}}{3H^2} \right)
+ \mathcal{O}\left(\xi^3\right) \right|_{aH=k}
\,.\end{eqnarray}

\newpage

\subsection{Power spectrum}

Now, to calculate the power spectrum, we substitute
Eq.~(\ref{gexp}) into Eq.~(\ref{gfsol}) and integrate iteratively.
Most of the calculations have already been done in \cite{sg}, and,
in the limit $x\rightarrow 0$, the asymptotic form for
$\varphi^\mathbf{a}$ up to second-order corrections is
\begin{equation}\label{asym}
\varphi^\mathbf{a}(\mathbf{k},x) \rightarrow \frac{1}{\sqrt{2k}\,}
\frac{i}{x} \left[ a^\mathbf{a}(\mathbf{k}) -
a^{\dagger\mathbf{a}}(\mathbf{-k}) + Z^\mathbf{a}_\mathbf{\
b}(k,x)\,a^\mathbf{b}(\mathbf{k}) - {Z^\mathbf{a}_\mathbf{\
b}}^*(k,x)\, a^{\dagger\mathbf{b}}(\mathbf{-k}) \right]
\,,\end{equation} where
\begin{eqnarray}
Z^\mathbf{ab} & = & \left( \alpha + \frac{i\pi}{2} \right)
{\zeta_1}^\mathbf{ab} + \frac{1}{2} \left[ \alpha^2 -
\frac{2}{3}\alpha - 4 + \frac{\pi^2}{4} + i\pi
\left(\alpha-\frac{1}{3}\right) \right]
{\zeta_1}^\mathbf{a}_\mathbf{\ c} {\zeta_1}^\mathbf{cb} \nonumber
\\ & & \mbox{} + \frac{1}{2} \left[ \alpha^2 + \frac{2}{3}\alpha -
\frac{\pi^2}{12} + i\pi \left(\alpha+\frac{1}{3}\right) \right]
{\zeta_2}^\mathbf{ab} \nonumber
\\ & & \mbox{} - \left[ {\zeta_1}^\mathbf{ab} + \left( \alpha
-\frac{1}{3} +\frac{i\pi}{2} \right)
{\zeta_1}^\mathbf{a}_\mathbf{\ c} {\zeta_1}^\mathbf{cb} +
\frac{1}{3} {\zeta_2}^\mathbf{ab} \right] \ln x \nonumber \\ & &
\mbox{} + \frac{1}{2} \left( {\zeta_1}^\mathbf{a}_\mathbf{\ c}
{\zeta_1}^\mathbf{cb} - {\zeta_2}^\mathbf{ab} \right) (\ln x)^2
\,.
\end{eqnarray}
The numerical constant $\alpha$ is defined as
\begin{equation}
\alpha \equiv 2 - \ln 2 - \gamma \simeq 0.729637
\,,\end{equation}
where $\gamma \simeq 0.577216$ is the Euler-Mascheroni constant.
Eq.~(\ref{asym}) gives
\begin{eqnarray}\label{vev}
\lefteqn{ \left\langle \varphi^\mathbf{a}(\mathbf{k},x)\,
{\varphi^\mathbf{b}}^\dagger(\mathbf{l},x) \right\rangle =
\frac{1}{2k x^2}\,\delta^{(3)}(\mathbf{k-l}) \left[ h^\mathbf{ab}
+ Z^\mathbf{ab} + {Z^\mathbf{ba}}^* + h_\mathbf{cd} Z^\mathbf{ac}
{Z^\mathbf{bd}}^* \right] } \nonumber \\ & = & \frac{1}{2k
x^2}\,\delta^{(3)}(\mathbf{k-l}) \times \nonumber \\ & & \left\{
h^\mathbf{ab} + 2\alpha {\zeta_1}^\mathbf{ab} +
\left(2\alpha^2-\frac{2}{3}\alpha-4+\frac{\pi^2}{2}\right)
{{\zeta_1}^\mathbf{a}}_\mathbf{c} {\zeta_1}^\mathbf{cb} + \left(
\alpha^2 + \frac{2}{3}\alpha - \frac{\pi^2}{12} \right)
{\zeta_2}^\mathbf{ab} \right. \nonumber \\ & & \left. \mbox{} - 2
\left[ {\zeta_1}^\mathbf{ab} + \left( 2\alpha - \frac{1}{3}
\right) {{\zeta_1}^\mathbf{a}}_\mathbf{c}{\zeta_1}^\mathbf{cb} +
\frac{1}{3} {\zeta_2}^\mathbf{ab} \right] \ln x + \left( 2
{\zeta_1}^\mathbf{a}_\mathbf{\ c} {\zeta_1}^\mathbf{cb} -
{\zeta_2}^\mathbf{ab} \right) (\ln x)^2 \right\}
\end{eqnarray}
and so, from Eq.~(\ref{p1}), the power spectrum is
\begin{eqnarray}\label{p2}
\mathcal{P}(k) & = & N_\mathbf{,a} N_\mathbf{,b} \left(
\frac{k}{2\pi ax} \right)^2 \times \nonumber \\ & & \left\{
h^\mathbf{ab} + 2\alpha {\zeta_1}^\mathbf{ab} +
\left(2\alpha^2-\frac{2}{3}\alpha-4+\frac{\pi^2}{2}\right)
{{\zeta_1}^\mathbf{a}}_\mathbf{c} {\zeta_1}^\mathbf{cb} + \left(
\alpha^2 + \frac{2}{3}\alpha - \frac{\pi^2}{12} \right)
{\zeta_2}^\mathbf{ab} \right. \nonumber \\ & & \left. \mbox{} - 2
\left[ {\zeta_1}^\mathbf{ab} + \left( 2\alpha - \frac{1}{3}
\right) {{\zeta_1}^\mathbf{a}}_\mathbf{c}{\zeta_1}^\mathbf{cb} +
\frac{1}{3} {\zeta_2}^\mathbf{ab} \right] \ln x + \left( 2
{\zeta_1}^\mathbf{a}_\mathbf{\ c} {\zeta_1}^\mathbf{cb} -
{\zeta_2}^\mathbf{ab} \right) (\ln x)^2 \right\} \,.
\end{eqnarray}

The right hand side is constant to order $\xi^2$ and so we can
choose to evaluate it at any convenient time around horizon
crossing. We will evaluate it at $aH=k$.

Now, from Eq.~(\ref{eta}),
\begin{equation}
x |_{aH=k} \simeq 1 + \epsilon + 3\epsilon^{\ 2}
+ 2\epsilon\delta |_{aH=k}
\end{equation}
Therefore
\begin{eqnarray}\label{pfinal}
\mathcal{P}(k) & = & \left( \frac{H}{2\pi} \right)^2 N_\mathbf{,a}
N_\mathbf{,b} \left\{ \left( 1 - 2\epsilon - 3\epsilon^{\ 2} -
4\epsilon\delta \right) h^\mathbf{ab} + \left[ 2\alpha -
(4\alpha+2)\epsilon \right] {\zeta_1}^\mathbf{ab} \right. \nonumber \\
& & \left. \left. \mbox{} + \left(2\alpha^2 - \frac{2}{3}\alpha -
4 + \frac{\pi^2}{2}\right) {\zeta_1^\mathbf{\ a}}_\mathbf{c}
{\zeta_1}^\mathbf{cb} + \left( \alpha^2 + \frac{2}{3}\alpha
-\frac{\pi^2}{12} \right) {\zeta_2}^\mathbf{ab} \right\}
\right|_{aH=k} \,,\end{eqnarray} where ${\zeta_1}^\mathbf{ab}$ and
${\zeta_2}^\mathbf{ab}$ are given by Eqs.~(\ref{g1})
and~(\ref{g2}).

To write $\mathcal P(k)$ in terms of the inflaton potential,
we define
\begin{equation}
\mathcal{U} \equiv \frac{V^\mathbf{,a}}{V}\frac{V_\mathbf{,a}}{V}
\,, \hspace{1em}
\mathcal{V}_\mathbf{ab} \equiv \frac{V_\mathbf{;ab}}{V}
\,, \hspace{1em}
\mathcal{V}_\mathbf{abc}
\equiv \mathcal{U}^\frac{1}{2} \frac{V_\mathbf{;abc}}{V}
\,,\end{equation}
the unit covectors
\begin{equation}
e^N_\mathbf{a} \equiv
\frac{N_\mathbf{,a}}{\sqrt{N^\mathbf{,b}N_\mathbf{,b}}}
\hspace{1em} \mbox{and} \hspace{1em}
e^V_\mathbf{a} \equiv
\frac{V_\mathbf{,a}}{\sqrt{V^\mathbf{,b}V_\mathbf{,b}}}
\,,\end{equation}
the unit vectors
\begin{equation}
e_N^\mathbf{a} \equiv h^\mathbf{ab} e^N_\mathbf{b}
\hspace{1em} \mbox{and} \hspace{1em}
e_V^\mathbf{a} \equiv h^\mathbf{ab} e^V_\mathbf{b}
\,,\end{equation}
and the component notation
\begin{equation}
w_N \equiv e_N^\mathbf{a} w_\mathbf{a}
\hspace{1em} \mbox{and} \hspace{1em}
w_V \equiv e_V^\mathbf{a} w_\mathbf{a}
\,.\end{equation}
Now
\begin{equation}
\frac{V_\mathbf{,a}}{V} = \mathcal{U}^\frac{1}{2} e^V_\mathbf{a}
\end{equation}
and using Eqs.~(\ref{bgeq}) and~(\ref{etceq}) we can derive
\begin{equation}
H^2 = \frac{V}{3} \left[
1 + \frac{1}{6} \mathcal{U} - \frac{1}{12} \mathcal{U}^2
+ \frac{1}{9} \mathcal{U} \mathcal{V}_{VV}
+ \mathcal{O}\left(\xi^3\right) \right]
\,,\end{equation}
\begin{equation}
\epsilon = \frac{1}{2} \mathcal{U} \left(
1 - \frac{2}{3} \mathcal{U} + \frac{2}{3} \mathcal{V}_{VV} \right)
+ \mathcal{O}\left(\xi^3\right)
\,,\end{equation}
\begin{equation}
\delta = \frac{1}{2} \mathcal{U} - \mathcal{V}_{VV}
+ \mathcal{O}\left(\xi^2\right)
\,,\end{equation}
\begin{equation}
\frac{\dot\phi_\mathbf{a}}{H}\frac{\dot\phi_\mathbf{b}}{H}
= \left( 1 - \frac{2}{3} \mathcal{U} \right)
\mathcal{U} e^V_\mathbf{a} e^V_\mathbf{b}
+ \frac{1}{3} \mathcal{U} \left(
e^V_\mathbf{a} \mathcal{V}_{\mathbf{b}V}
+ e^V_\mathbf{b} \mathcal{V}_{\mathbf{a}V} \right)
+ \mathcal{O}\left(\xi^3\right)
\,,\end{equation}
\begin{eqnarray}
\frac{D}{H dt} =
- \mathcal{U}^\frac{1}{2} \nabla_V
- \frac{1}{3} \mathcal{U}^\frac{1}{2}
h^\mathbf{ab} \mathcal{V}_{\mathbf{a}V} \nabla_\mathbf{b}
+ \mathcal{O}\left(\xi^2\right)
\,.\label{del}\end{eqnarray}
The second term in Eq.~(\ref{del}) is needed because, for example,
$\mathcal{V}_{\mathbf{a}V\mathbf{b}}
= \mathcal{V}_{\mathbf{ab}V}
- \mathcal{U} R_{\mathbf{a}VV\mathbf{b}}$ and
$\mathcal{U} R_{\mathbf{a}VV\mathbf{b}}$ is of order $\xi$.
Thus, while $\mathcal{V}_{\mathbf{ab}V}$ is of order $\xi^2$,
$\mathcal{V}_{\mathbf{a}V\mathbf{b}}$ is of order $\xi$.
We could drop the second term in Eq.~(\ref{del})
if we assumed either $\mathcal{U} = \mathcal{O}(\xi^2)$,
which is physically very reasonable but non-standard,
or $R_{\mathbf{a}VV\mathbf{b}} = \mathcal{O}(\xi)$,
which is possible but not what one would expect in general.

Expressing Eqs.~(\ref{pfinal}), (\ref{g1}) and~(\ref{g2})
in terms of the potential gives
\begin{eqnarray}\label{semifinal}
\mathcal{P}(k) & = & \frac{V N_\mathbf{,e}N^\mathbf{,e}}{12\pi^2}
\left\{ 1 - \frac{5}{6}\mathcal{U} - \frac{4}{3}\mathcal{U}^2 +
\frac{13}{9} \mathcal{U} \mathcal{V}_{VV} +
\left[2\alpha-\left(\frac{5}{3}\alpha+1\right)\mathcal{U}\right]
{\zeta_1}_{NN} \right. \nonumber \\ & & \left. \mbox{} +
\left(2\alpha^2-\frac{2}{3}\alpha-4+\frac{\pi^2}{2}\right)
h^\mathbf{ab} {\zeta_1}_{N\mathbf{a}} {\zeta_1}_{\mathbf{b}N} +
\left( \alpha^2 + \frac{2}{3}\alpha - \frac{\pi^2}{12} \right)
{\zeta_2}_{NN} \right\} \,,\end{eqnarray} where
\begin{eqnarray}
{\zeta_1}_\mathbf{ab} & = & \frac{1}{2} \left( 1 +
\frac{8}{3}\mathcal{U} - 2 \mathcal{V}_{VV} \right) \mathcal{U}
h_\mathbf{ab} + \left( 1 + \frac{5}{6}\mathcal{U} \right) \left(
\mathcal{U} e^V_\mathbf{a} e^V_\mathbf{b} -
\mathcal{V}_\mathbf{ab} \right) \nonumber \\ & & \mbox{} +
\frac{1}{3} \left( 1 + \frac{1}{3}\mathcal{U} \right) \mathcal{U}
R_{\mathbf{a}VV\mathbf{b}} + \frac{1}{9} \mathcal{U} h^\mathbf{cd}
\left( R_{\mathbf{a}V\mathbf{cb}} + R_{\mathbf{ac}V\mathbf{b}}
\right) \mathcal{V}_{\mathbf{d}V} \label{g1V}\end{eqnarray} and
\begin{eqnarray}
{\zeta_2}_\mathbf{ab} & = & \mathcal{U} \left( \mathcal{V}_{VV} -
\mathcal{U} \right) h_\mathbf{ab} - 2 \mathcal{U}^2 e^V_\mathbf{a}
e^V_\mathbf{b} + \mathcal{U} \mathcal{V}_\mathbf{ab} + \mathcal{U}
\left( e^V_\mathbf{a} \mathcal{V}_{\mathbf{b}V} + e^V_\mathbf{b}
\mathcal{V}_{\mathbf{a}V} \right) - \mathcal{V}_{\mathbf{ab}V} -
\frac{1}{3} h^\mathbf{cd} \mathcal{V}_{\mathbf{abc}}
\mathcal{V}_{\mathbf{d}V} \nonumber \\ & & \mbox{} - \frac{2}{3}
\mathcal{U}^2 R_{\mathbf{a}VV\mathbf{b}} + \frac{1}{3} \mathcal{U}
h^\mathbf{cd} \left( R_{\mathbf{a}V\mathbf{cb}} +
R_{\mathbf{ac}V\mathbf{b}} \right) \mathcal{V}_{\mathbf{d}V} +
\frac{1}{3} \mathcal{U}^\frac{3}{2} \left(
R_{\mathbf{a}VV\mathbf{b};V} + \frac{1}{3} h^\mathbf{cd}
R_{\mathbf{a}VV\mathbf{b};\mathbf{c}} \mathcal{V}_{\mathbf{d}V}
\right) \,. \nonumber \\ \label{g2V}\end{eqnarray} Substituting
Eqs.~(\ref{g1V}) and~(\ref{g2V}) into Eq.~(\ref{semifinal}), we
can obtain the power spectrum in terms of the inflaton potential
and its derivatives explicitly. The result is
\begin{eqnarray}\label{p-potential}
\mathcal{P}(k) & = & \frac{V N_\mathbf{,c} N^\mathbf{,c}}{12\pi^2}
\left\{
1 + \left( \alpha - \frac{5}{6} \right)\mathcal U
+ \left(-\frac{1}{2}\alpha^2+\alpha-\frac{17}{6}+\frac{5\pi^2}{24} \right) \mathcal U^2
\right. \nonumber \\ & & \hspace{5em} \mbox{}
+ \left[ 2\alpha + \left( 2\alpha^2 - \frac{8}{3}\alpha - 9 + \frac{7\pi^2}{6} \right) \mathcal{U} \right] \mathcal{U} h_{NV}^2
\nonumber \\ & & \hspace{5em} \mbox{}
+ \left[ - 2\alpha\mathcal + \left( -\alpha^2 + \frac{4}{3}\alpha + 5 - \frac{7\pi^2}{12} \right) \mathcal{U} \right]
\mathcal{V}_{NN}
\nonumber \\ & & \hspace{5em} \mbox{}
+ \left[ \frac{2}{3} \alpha + \left( - \alpha - \frac{5}{3} + \frac{2\pi^2}{9} \right) \mathcal{U} \right]
\mathcal{U} R_{NVVN}
\nonumber \\ & & \hspace{5em} \mbox{}
+ \left(\alpha^2-\frac{4}{3}\alpha+\frac{13}{9}-\frac{\pi^2}{12} \right) \mathcal{U} \mathcal{V}_{VV}
\nonumber \\ & & \hspace{5em} \mbox{}
+ \left(-2\alpha^2+\frac{8}{3}\alpha+8-\frac{7\pi^2}{6}\right)
\mathcal{U} h_{NV} \mathcal{V}_{NV}
\nonumber \\ & & \hspace{5em} \mbox{}
+ \left( -\alpha^2 - \frac{2}{3}\alpha + \frac{\pi^2}{12} \right) \left( \mathcal{V}_{NNV} + \frac{1}{3} h^\mathbf{ab}
\mathcal{V}_{NN\mathbf{a}} \mathcal{V}_{\mathbf{b}V} \right)
\nonumber \\ & & \hspace{5em} \mbox{}
+ \left(\frac{1}{3}\alpha^2+\frac{2}{9}\alpha-\frac{\pi^2}{36}
\right) \mathcal{U}^\frac{3}{2}
\left( R_{NVVN;V} + \frac{1}{3} h^\mathbf{ab} R_{NVVN;\mathbf{a}}
\mathcal{V}_{\mathbf{b}V} \right)
\nonumber \\ & & \hspace{5em} \mbox{}
+ \left(2\alpha^2-\frac{2}{3}\alpha-4+\frac{\pi^2}{2}\right)
h^\mathbf{ab} \left( \mathcal{V}_{\mathbf{a}N}
- \frac{1}{3} \mathcal{U} R_{NVV\mathbf{a}} \right)
\left( \mathcal{V}_{\mathbf{b}N}
- \frac{1}{3} \mathcal{U} R_{NVV\mathbf{b}} \right)
\nonumber \\ & & \hspace{5em} \left. \mbox{}
+ \left( \frac{2}{3}\alpha^2+\frac{8}{9}\alpha-\frac{\pi^2}{18}
\right) \mathcal{U}
h^\mathbf{ab} R_{NV\mathbf{a}N} \mathcal{V}_{\mathbf{b}V}
+ \mathcal{O}\left(\xi^3\right)
\right\}
\,.\end{eqnarray}

Finally, the spectral index,
$n-1 \equiv d\ln\mathcal{P} / d\ln k$,
and the running of the spectral index can be calculated from Eq.~(\ref{p2}) in a similar way.
The results are
\begin{eqnarray}
n - 1 & = &
-\mathcal{U}
+ \left( 2\alpha - \frac{11}{6} \right) \mathcal{U}^2
+ \left( -2 + \mathcal{U} \right) \mathcal{U}h_{NV}^2
\nonumber \\ & & \mbox{}
+ \left[ 2 + \left( -2\alpha + \frac{1}{3} \right)\mathcal{U}
\right] \mathcal{V}_{NN}
+ \left[ -\frac{2}{3} + \left( \frac{4}{3}\alpha + \frac{4}{9} \right) \mathcal{U} \right] \mathcal{U}R_{NVVN}
\nonumber \\ & & \mbox{}
+ \left(-2\alpha+\frac{4}{3}\right) \mathcal{U}\mathcal{V}_{VV}
+ \left( 4\alpha - \frac{8}{3} \right) \mathcal{U} h_{NV} \mathcal{V}_{NV}
\nonumber \\ & & \mbox{}
+ 4\alpha \left( \mathcal{U}h_{NV}^2 - \mathcal{V}_{NN} + \frac{1}{3}\mathcal{U}R_{NVVN} \right)^2
\nonumber \\ & & \mbox{}
+ \left( 2\alpha + \frac{2}{3} \right)
\left( \mathcal{V}_{NNV} + \frac{1}{3} h^\mathbf{ab} \mathcal{V}_{NN\bf a} \mathcal{V}_{\mathbf{b}V} \right)
\nonumber \\ & & \mbox{}
+ \left( - \frac{2}{3}\alpha - \frac{2}{9} \right) \mathcal{U}^\frac{3}{2} \left( R_{NVVN;V}
+ \frac{1}{3} h^\mathbf{ab} R_{NVVN;\mathbf{a}} \mathcal{V}_{\mathbf{ b}V} \right)
\nonumber \\ & & \mbox{}
+ \left( - 4\alpha + \frac{2}{3} \right) h^\mathbf{ab}
\left( \mathcal{V}_{\mathbf{a}N} - \frac{1}{3}\mathcal{U}R_{NVV\mathbf{a}} \right)
\left(\mathcal{V}_{\mathbf{b}N} - \frac{1}{3}\mathcal{U}R_{NVV\mathbf{ b}} \right)
\nonumber \\ & & \mbox{}
+ \left( - \frac{4}{3}\alpha - \frac{8}{9} \right) \mathcal{U} h^\mathbf{ab} R_{NV\mathbf{a}N} \mathcal{V}_{\mathbf{b}V}
+ \mathcal{O}\left(\xi^3\right)
\end{eqnarray}
and
\begin{eqnarray}
\frac{d n}{d \ln k} & = & - 2 \mathcal{U}^2 + 2 \mathcal{U}
\mathcal{V}_{NN} - \frac{4}{3} \mathcal{U}^2 R_{NVVN} + 2
\mathcal{U} \mathcal{V}_{VV} - 4 \mathcal{U} h_{NV}
\mathcal{V}_{NV} \nonumber \\ & & \mbox{} - 4 \left( \mathcal{U}
h_{NV}^2 - \mathcal{V}_{NN} + \frac{1}{3} \mathcal{U} R_{NVVN}
\right)^2 \nonumber \\ & & \mbox{} - 2 \left( \mathcal{V}_{NNV} +
\frac{1}{3} h^\mathbf{ab} \mathcal{V}_{NN\mathbf{a}}
\mathcal{V}_{\mathbf{b}V} \right) \nonumber \\ & & \mbox{} +
\frac{2}{3} \mathcal{U}^\frac{3}{2} \left( R_{NVVN;V} +
\frac{1}{3} h^\mathbf{ab} R_{NVVN;\mathbf{a}}
\mathcal{V}_{\mathbf{b}V} \right) \nonumber \\ & & \mbox{} + 4
h^\mathbf{ab} \left( \mathcal{V}_{\mathbf{a}N} - \frac{1}{3}
\mathcal{U} R_{NVV\mathbf{a}} \right) \left(
\mathcal{V}_{\mathbf{b}N} - \frac{1}{3} \mathcal{U}
R_{NVV\mathbf{b}} \right) \nonumber \\ & & \mbox{} + \frac{4}{3}
\mathcal{U} h^\mathbf{ab} R_{NV\mathbf{a}N}
\mathcal{V}_{\mathbf{b}V} + \mathcal{O}\left(\xi^3\right).
\end{eqnarray}

\subsection*{Acknowledgements}

We thank Hyun-Chul Lee for valuable discussions and helpful
comments. This work was supported in part by Brain Korea 21 and
KRF grant 2000-015-DP0080.

\end{document}